\documentclass[10pt,journal,twocolumn]{IEEEtran}

\usepackage{amsmath}
\usepackage{graphicx}
\usepackage{algorithm}
\usepackage{algpseudocode}
\usepackage{booktabs}
\usepackage{multirow}
\usepackage{cite} 

\begin{document}

\title{Taming the Black Swan: A Momentum-Gated Hierarchical Optimisation Framework for Asymmetric Alpha Generation}

\author{
    \IEEEauthorblockN{Arya Chakraborty\IEEEauthorrefmark{1}, Dr. Randhir Singh\IEEEauthorrefmark{2}} \\
    \IEEEauthorblockA{\textit{Department of Computer Science\IEEEauthorrefmark{1}, Department of Mathematics\IEEEauthorrefmark{2}} \\
    \textit{Birla Institute of Technology Mesra, Ranchi, India} \\
    btech10196.23@bitmesra.ac.in\IEEEauthorrefmark{1}, aryachakraborty2005@gmail.com\IEEEauthorrefmark{1}, randhir.math@gmail.com\IEEEauthorrefmark{2}}
}

\maketitle

\begin{abstract}
Conventional momentum strategies, despite their proven efficacy in generating alpha, frequently suffer from the \textit{``Winner's Curse''}, a structural vulnerability in which high-performing assets exhibit clustered volatility and severe drawdowns during market reversals. To counteract this propensity for momentum crashes, this study presents the \textit{Adaptive Equity Generation and Immunisation System} (AEGIS), a novel framework that fundamentally reengineers the trade-off between growth and stability. By leveraging a volatility-adjusted momentum filter to identify trend strength and employing a \textit{minimax correlation} algorithm to enforce structural diversification, the model utilises \textit{sequential least squares programming} (SLSQP) to optimise capital allocation for the sortino ratio. This architecture allows the portfolio to dynamically adapt to distinct market regimes: explicitly lowering the intensity of crashes during bear markets by \textit{decoupling correlated risks}, while \textit{retaining asymmetric upside participation} during bull runs. Empirical validation via a comprehensive 20-year walk-forward backtest (2006--2025), which covers significant stress events like the 2008 Global Financial Crisis, confirms that the framework produces substantial excess alpha relative to the standard S\&P 500 benchmark. Notably, the strategy successfully matched the capital appreciation of the high-beta NASDAQ-100 index while achieving significantly reduced downside volatility and improved structural resilience. These results suggest that synthetic beta can be effectively engineered through mathematical regularisation, enabling investors to capture the high-growth characteristics of concentrated portfolios while preserving the defensive stability typically associated with broad-market diversification.
\end{abstract}

\begin{IEEEkeywords}
Portfolio Optimisation, Minimax Correlation, Momentum Investing, SLSQP (Sequential Least Squares Programming), Sortino Ratio, Downside Risk Management, Algorithmic Trading
\end{IEEEkeywords}

\section{Introduction}
Momentum investing is the strategy of buying past winners and selling past losers. It stands out as one of the most robust and empirically validated anomalies in modern finance \cite{ref1, ref18, ref21}. This technique challenges the \textit{Efficient Market Hypothesis} (EMH) \cite{ref2} by demonstrating that asset prices exhibit significant persistence over intermediate horizons, typically three to twelve months. The tendency of high-performing assets to continue outperforming and generate substantial excess returns across diverse asset classes, geographies, and time periods has made it a cornerstone of quantitative equity strategies. The persistence of momentum is frequently attributed to behavioural biases among market participants \cite{ref14, ref16, ref17}. The underreaction-overreaction mechanism suggests that investors initially underreact to news due to conservatism, only to later chase returns, driving prices beyond fundamental value \cite{ref3, ref15}. Alternative risk-based explanations argue that momentum profits are simply compensation for exposure to time-varying systematic risks \cite{ref4}. Regardless, the strategy’s ability to generate alpha in standard market conditions is well-documented. However, the excess return generated comes at a distinct structural cost. Momentum strategies are notoriously susceptible to momentum crashes, which are sudden, catastrophic reversals that occur when the market transitions from a downtrend to a sharp rebound \cite{ref5, ref59}. During these stress events, the correlation among winning assets spikes, leading to clustered volatility. This phenomenon, often termed the \textit{``Winner's Curse''} reveals that standard momentum portfolios effectively hold a short position in market volatility. Since high-momentum assets often carry high betas or extreme valuations, they suffer disproportionately when market sentiment shifts, leading to significant drawdowns that can erase years of accumulated gains in a matter of months \cite{ref5}. 

To mitigate these structural risks, modern quantitative practitioners typically rely on \textit{Modern Portfolio Theory} and \textit{Mean-Variance Optimisation} (MVO) to construct diversified Portfolios \cite{ref6, ref24}. Pioneered by Markowitz, this framework seeks to maximise returns for a given level of risk by diversifying across assets with low covariance. While mathematically elegant, MVO relies on the gaussian assumption that asset returns follow a normal distribution, a premise that frequently disintegrates during the fat-tailed events characteristic of momentum crashes. Crucially, conventional mean-variance models fail to distinguish between `good' upside volatility and `bad' downside volatility \cite{ref7, ref27}. By penalising the total variance indiscriminately, these frameworks often cap the potential alpha of high-momentum assets, effectively forcing the strategy to exit winners precisely when their trend is strongest \cite{ref7}. Furthermore, standard diversification often creates a false sense of security; during liquidity crises, correlations across all asset classes tend to converge to 1, rendering traditional covariance matrices useless when they are needed most \cite{ref8, ref35, ref40}. This creates a critical dilemma: investors are forced to choose between the unmitigated tail risk of concentrated momentum or the diluted returns of over-diversified mean-variance portfolios. This theoretical inadequacy is most pronounced during \textit{`Black Swan'} events, which are rare, unpredictable outliers carrying extreme impact, yet being retrospectively explainable \cite{ref9}. Traditional optimisation models, predicated on the Gaussian bell curve, treat such market dislocations as statistical impossibilities. However, empirical evidence shows that asset returns exhibit significant leptokurtosis (fat tails) and negative skewness, indicating that extreme loss events occur far more frequently than normal distributions predict \cite{ref35, ref36}. In this non-linear environment, the primary threat to long-term wealth accumulation is not daily volatility, but the absorbing barrier of ruin, i.e., a depth of drawdown from which recovery becomes mathematically improbable \cite{ref10}. Consequently, a robust equity framework must move beyond simple mean-variance trade-offs to explicitly target tail-risk immunisation, prioritising survivability during systemic shocks without sacrificing the compounding power of the momentum anomaly. To navigate the non-linear risk landscape of modern markets, this study introduces the \textit{Adaptive Equity Generation and Immunisation System} (AEGIS). Unlike traditional mean-variance frameworks that penalise total volatility indiscriminately, thereby capping potential alpha by treating upside surges as risk, AEGIS operates on a three-layered architecture that is designed to fundamentally decouple return generation from tail-risk exposure. The framework acknowledges that in a fat-tailed world, the primary objective of portfolio construction must shift from simple variance reduction to the explicit maximisation of the survival function. By synthesising a volatility-adjusted momentum filter with a robust correlation screening mechanism, the system moves beyond naive diversification, which often fails when correlations converge to one during liquidity crises. Instead, AEGIS dynamically engineers a basket of high-efficiency trend followers (the anchors) that are mathematically constrained to maintain structural independence, ensuring that the portfolio does not merely accumulate correlated winners but rather builds a resilient, multi-sector alpha engine. This is followed by selecting noncorrelated diversifiers to capture upside and reduce potential drawdown during high-stress events. This approach allows targeting a convex payoff profile that captures asymmetric upside participation while imposing a strict barrier against catastrophic drawdowns. The efficacy of this philosophy is borne out by a comprehensive 20-year walk-forward analysis covering multiple market regimes, including severe stress events such as the \textit{2008 Financial Crisis} and the \textit{2022 Inflation Shock}. Qualitatively, the results demonstrate a significant decoupling from broad market benchmarks. The AEGIS framework generated substantial excess returns relative to the S\&P 500, effectively delivering a multiple of the market's total wealth accumulation. More notably, the strategy successfully matched the aggressive capital appreciation of the high-beta NASDAQ-100 index but achieved this growth trajectory with a drastically superior risk profile. Therefore, AEGIS offers a blueprint for a new class of anti-fragile portfolios, demonstrating that survivability in the face of black swan events is not a function of prediction, but of structural design.

\begin{table}[htbp]
\centering
\caption{List of Key Abbreviations}
\label{tab:abbreviations}
\begin{tabular}{@{}ll@{}}
\toprule
\textbf{Abbreviation} & \textbf{Definition} \\ \midrule
AEGIS & Adaptive Equity Generation and Immunisation System \\
CAGR  & Compound Annual Growth Rate \\
CSM   & Cross-Sectional Momentum \\
DD    & Downside Deviation \\
GAAP   & Generally Accepted Accounting Principles \\
FALR  & Float-Adjusted Liquidity Ratio \\
ADVT  & Annual Dollar Value Traded \\
FMC   & Float-Adjusted Market Capitalization \\
GFC   & Global Financial Crisis \\
GICS  & Global Industry Classification Standard \\
IWF   & Investable Weight Factor \\
LPM   & Lower Partial Moment \\
MAR   & Minimum Acceptable Return \\
MDD   & Maximum Drawdown \\
MPT   & Modern Portfolio Theory \\
MVO   & Mean-Variance Optimisation \\
SLSQP & Sequential Least Squares Programming \\
TR    & Total Return \\
VAM   & Volatility-Adjusted Momentum \\ \bottomrule
\end{tabular}
\end{table}

\section{Methodology}

\begin{figure*}[htbp]
\centering
\includegraphics[width=\textwidth]{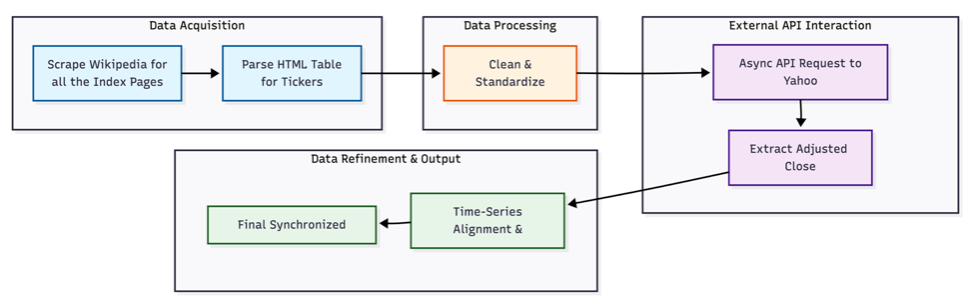} 
\caption{Automated Data Acquisition Architecture. The pipeline scrapes constituent tickers from Wikipedia, executes asynchronous requests to the Yahoo Finance API for adjusted close prices, and performs time-series alignment to generate a synchronised historical dataset for analysis.}
\label{fig:arch1}
\end{figure*}

\subsection{Data Acquisition and Processing}
The investment universe is constructed by aggregating constituents from five primary U.S. equity indices: the S\&P 500, S\&P MidCap 400, S\&P SmallCap 600, NASDAQ-100, and the Dow Jones Industrial Average. These benchmarks were selected for their rigorous, rule-based inclusion criteria, which inherently filter for liquidity, financial viability, and structural quality \cite{ref50, ref51}. The S\&P suite (500, 400, 600) employs a unified selection methodology managed by the index committee. Candidates must meet strict quantitative thresholds to ensure investability and financial health. \vspace{0.25cm}

\subsubsection{S\&P Suite Selection Protocol}
The S\&P Suite presents a particularly interesting structured approach to the selection of assets for all of its indices.

\noindent \textbf{Market Capitalisation Thresholds:}
The precise unadjusted market capitalisation is the total market value of a company’s outstanding shares, calculated by multiplying the current share price by the total number of shares issued, without accounting for free-float factors such as restricted stock or shares held by insiders.
\begin{equation}
\text{Unadjusted Cap} = \text{Share Price} \times \text{Shares Outstanding}.
\end{equation}

To classify the universe by size, precise unadjusted market capitalisation ($MC_{unadj}$) bands are enforced:
\begin{itemize}
    \item[(1)] S\&P 500 (Large-Cap): $MC_{unadj} \ge \$22.7 \text{ Billion}$.
    \item [(2)] MidCap 400: $\$8.0 \text{ Billion} \le MC_{unadj} < \$22.7 \text{ Billion}$.
    \item [(3)] SmallCap 600: $\$1.2 \text{ Billion} \le MC_{unadj} < \$8.0 \text{ Billion}$.
\end{itemize}

\noindent \textbf{Liquidity Requirement:}
The Float-Adjusted Liquidity Ratio is a metric that measures a stock's liquidity relative to the shares actually available to the public.
\begin{equation}
FALR = \frac{\text{ADVT}}{\text{FMC}},
\end{equation}
where ADVT is the average closing price multiplied by the total volume of shares traded over the last 365 days, and FMC is the market value of only the shares available to the public (excluding shares held by insiders, founders, or governments). To ensure tradability, constituents must possess an FALR of $\ge 0.75$. Additionally, the stock must exhibit a minimum trading volume of 250,000 shares in each of the six months prior to evaluation.

\noindent \textbf{Float Adjustment (IWF):} The investable weight factor is a numerical value (ranging from 0 to 1) that represents the proportion of a company's total outstanding shares that are available for purchase by the public. The indices utilise IWF to isolate shares available to public investors. Only companies with an $IWF \ge 0.10$ are eligible. The weight $W_i$ of each constituent $i$ in the index is calculated as:
\begin{equation}
W_i = \frac{P_i \times S_{i} \times (IWF)_i}{D_{SP}},
\end{equation}
where $P_i$ is the price, $S_{i}$ is total shares outstanding, $(IWF)_i$ is the investable weight factor and $D_{SP}$ is the proprietary S\&P Divisor used to normalise the index value.

\noindent \textbf{Financial Viability:}
Unique to S\&P indices, candidates must demonstrate positive GAAP net income for the most recent quarter and a positive sum of GAAP net income over the preceding four consecutive quarters.
\begin{equation}
\sum_{q=1}^{4} (NI)_q > 0,
\end{equation}
where $(NI)_q$ is the net income of the $q^{th}$ quarter. If a company is moving within the S\&P 1500 (e.g., promoting from MidCap 400 to S\&P 500), it does not need to re-qualify for the earnings viability rule. Its tenure in the MidCap 400 counts as proof of quality. To reduce turnover, a current constituent is not automatically removed if its market cap dips slightly below the specified band limit, provided it stays within a reasonable rank range relative to the market peer group. \vspace{0.25cm}

\subsubsection{NASDAQ Selection Protocol}
The NASDAQ-100 selects the 100 largest non-financial companies listed on the NASDAQ exchange. Unlike the S\&P, it has no earnings requirement, allowing for the inclusion of high-growth firms operating at a loss. Companies ranked 1-75 by market capitalisation are automatically selected. Companies ranked 76-100 are selected only if they were existing members; otherwise, the highest-ranked non-members fill the remaining slots.

\noindent \textbf{Allocation:} To prevent concentration risk, the index employs a modified market cap weighting scheme. 

\noindent Stage 1: No single stock's weight ($w_i$) may exceed 24\%.
\begin{equation}
w_i \le 0.24.
\end{equation}

\noindent Stage 2: The collective weight of all constituents with individual weights exceeding 4.5\% must not exceed 48\% of the total index at any point.
\begin{equation}
\sum_{i \in \{w>0.045\}} w_i \le 0.48.
\end{equation}

Even when weights are capped, the methodology ensures rank order preservation. The largest company remains the largest by weight, even if its absolute influence is reduced. If a company is acquired or delisted, it is replaced immediately. The replacement is typically the highest-ranked non-member from the list of eligible companies, ensuring the index is always full at 100 members. \vspace{0.25cm}

\subsubsection{Dow Jones Selection Protocol}
The DJIA consists of 30 blue-chip companies selected qualitatively for their reputation and sustained growth. Unlike market-cap indices, the DJIA's weights are determined solely by share price. The index value is calculated as:
\begin{equation}
DJIA = \frac{1}{D_{DJ}}\sum_{i=1}^{30} P_i,
\end{equation}
where $D_{DJ}$ is the dow vivisor, which is adjusted for stock splits to ensure index continuity. This methodology structurally overweights companies with high nominal share prices regardless of their actual market valuation. Whenever a company executes a stock split (e.g., 2-for-1), the divisor is mathematically reduced to prevent the DJIA index from dropping overnight. This methodology naturally underweights high-flying growth stocks (which often split their shares to keep prices accessible) and overweights established, slower-growing industrial/financial firms (which often maintain high nominal share prices). This gives the Dow a permanent value and low volatility tilt compared to the S\&P 500 or NASDAQ. \vspace{0.25cm}

\begin{figure*}[htbp]
\centering
\includegraphics[width=\textwidth]{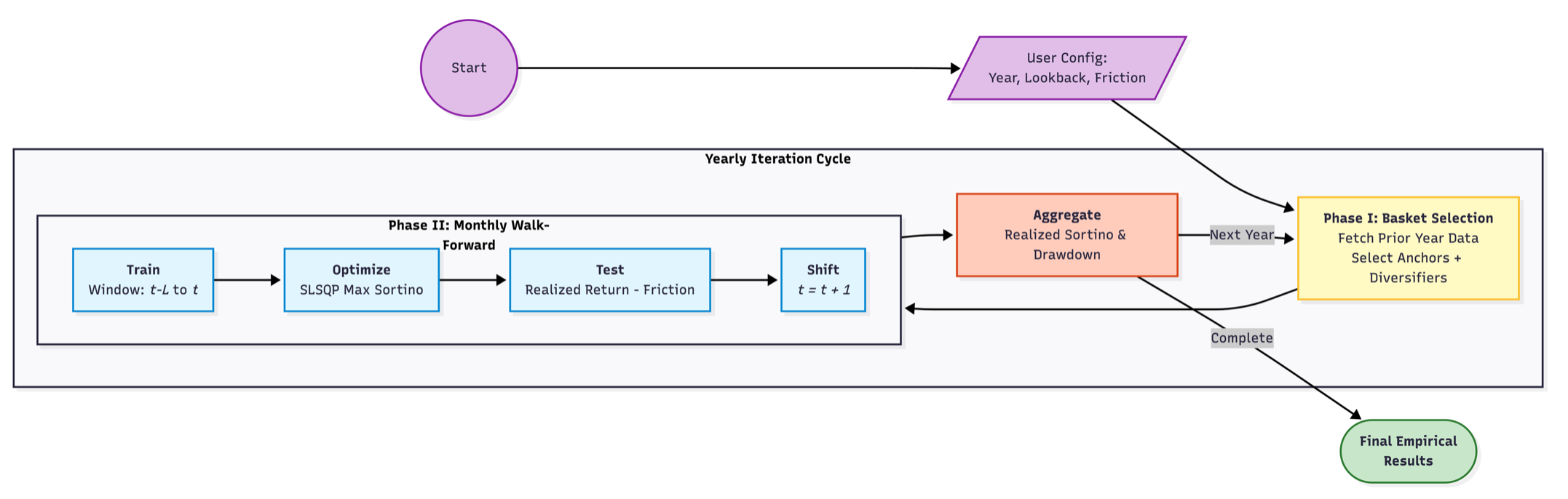} 
\caption{The AEGIS Walk-Forward Validation Architecture. The system operates via a nested-loop protocol: the outer annual cycle executes the Adaptive Basket Selection using prior-year momentum leaders, while the inner monthly cycle performs Rolling Optimisation. The framework strictly segregates the training window ($t - L \rightarrow t$) from the testing window ($t \rightarrow t + 1$) to eliminate look-ahead bias, generating realised performance metrics net of friction costs.}
\label{fig:arch2}
\end{figure*}

\subsubsection{Data Preprocessing}
To ensure the reproducibility of the experiment and the integrity of the input vectors, a robust, automated data pipeline was engineered. The system architecture, illustrated in Figure 1, operates as a multi-stage sequential workflow designed to handle high-volume financial time-series data with minimal latency.

\noindent \textbf{Constituent Aggregation:} The process initiates with the automated identification of the investment universe. A custom module was created to parse HTML tables from public repositories (specifically wikipedia’s constituent lists) to aggregate the ticker symbols for all five target indices (S\&P 500, S\&P 400, S\&P 600, NASDAQ-100, and Dow Jones). This module extracts the raw ticker strings and standardises them into a machine-readable format (e.g., BRK.B to BRK-B) for API compatibility.

\noindent \textbf{Concurrent API Information Retrieval:} To overcome the I/O bottlenecks inherent in fetching daily price history for thousands of assets over a multi-year horizon, the system utilises a concurrent request architecture.

\noindent Protocol: A ThreadPoolExecutor with 20 concurrent workers was employed to execute parallel HTTP requests to the \textit{Yahoo Finance} chart endpoint \cite{ref49, ref52}.

\noindent Data Extraction: The pipeline specifically retrieves the adjusted close price vector, which pre-adjusts for stock splits and dividend distributions, ensuring the dataset reflects $TR$ rather than nominal price appreciation.

\noindent Validation: Assets with insufficient history (fewer than 20 data points) or corrupted metadata are automatically flagged and rejected during ingestion.

\noindent \textbf{Time-Series Synchronisation:} Raw data retrieved from external APIs often contains gaps due to non-trading days, delistings, or API failures.

\noindent Alignment: All individual asset vectors are merged into a single $N \times T$ matrix (where $N$ is the number of assets and $T$ is the number of time steps) utilising an outer join on the date index.

\noindent Imputation: Missing values resulting from non-synchronous trading hours or data outages are handled using forward filling. This propagates the last known valid price forward, ensuring the optimiser never acts on future information (thereby eliminating look-ahead bias) while maintaining data continuity. 

\noindent Cleaning: Any remaining NaN rows (representing periods before an asset's IPO, after delisting, or when the data was not available from the API) are dropped to ensure a dense matrix for the correlation engine.

\subsection{Mathematical Framework}
To ensure the reproducibility of the experimental results, the specific mathematical formulations governing signal generation, risk filtration, and capital allocation are defined below. These components collectively form the theoretical basis of the AEGIS architecture. \vspace{0.25cm}

\subsubsection{The Momentum Factors}
The alpha generation engine relies on a sequential calculation of logarithmic returns, realised volatility, and a unified efficiency score.

\noindent \textbf{Logarithmic and Cumulative Returns:} To model continuous compounding and ensure time-additivity, the framework utilises logarithmic returns rather than simple arithmetic returns. The daily log-return $r_{i,t}$ for asset $i$ at time $t$ is defined as:
\begin{equation}
r_{i,t} = \ln\left(\frac{P_{i,t}}{P_{i,t-1}}\right),
\end{equation}
where $P_{i,t}$ is the adjusted closing price of asset $i$ at time $t$ and $P_{i,t-1}$ is the adjusted closing price of asset $i$ at time $t-1$. The cumulative return ($R_i$) over the look-back window $L$ (default $L=12$ months, approx. 252 trading days) is the summation of daily log-returns. To account for short-term microstructure reversals, the most recent month ($t-1$) is excluded (the skip-month method):
\begin{equation}
R_{i,t-L:t-1} = \sum_{k=t-L}^{t-21}r_{i,k} = \ln\left(\frac{P_{i,t-21}}{P_{i,t-L}}\right).
\end{equation}

\noindent \textbf{Realised Volatility:} Risk is quantified as the annualised standard deviation of daily log-returns over the same look-back window. This measures the dispersion of the asset's price path:
\begin{equation}
\sigma_i = \sqrt{\frac{252}{N-1}\sum_{k=1}^{N}(r_{i,k}-\bar{r}_i)^2},
\end{equation}
where $N$ is the total number of trading days in the look-back window, $\bar{r}_i$ is the mean daily log-return over the window, and 252 is the annual factor (approximate trading days in a year).

\noindent \textbf{Volatility-Adjusted Momentum:} Standard momentum strategies (ranking by $R_i$) often prioritise high-beta assets that exhibit extreme volatility. To mitigate this, AEGIS employs the VAM metric, which normalises return by risk \cite{ref60, ref61}, mathematically equivalent to a localised sharpe ratio:
\begin{equation}
S_{i,t} = \frac{R_{i,t-L:t-1}}{\sigma_i}.
\end{equation}
This metric prioritises assets with smooth, monotonic price appreciation over those with jagged, speculative spikes. \vspace{0.25cm}

\subsubsection{The Minimax Correlation}
To strictly enforce structural diversification, the framework employs a minimax correlation filter. This algorithm prevents risk clustering by minimising the worst-case dependency between assets \cite{ref23}.

\noindent \textbf{Pearson Correlation Coefficient} The dependency between any two assets $i$ and $j$ is quantified using the pearson product-moment correlation coefficient $\rho_{i,j}$:
\begin{equation}
\rho_{i,j} = \frac{\sum(r_{i,t}-\bar{r}_i)(r_{j,t}-\bar{r}_j)}{\sqrt{\sum(r_{i,t}-\bar{r}_i)^2}\sqrt{\sum(r_{j,t}-\bar{r}_j)^2}}.
\end{equation}
The pearson correlation coefficient ($r$) is a statistical measure that quantifies the strength and direction of a linear relationship between two continuous variables. It produces a value between -1.0 (perfect negative correlation) and +1.0 (perfect positive correlation), where 0 indicates no linear relationship exists.

\noindent \textbf{The Minimax Objective:} Unlike MVO, which minimises average portfolio variance, the AEGIS immunisation layer solves for a subset of assets $\Omega$ that minimises the maximum pairwise correlation. The optimisation problem is defined as:
\begin{equation}
\min_{\Omega} \left( \max_{i,j\in\Omega,i\neq j} |\rho_{i,j}| \right).
\end{equation}
This constraint ensures that no two assets in the portfolio share a correlation exceeding the minimax threshold, mathematically forcing the basket to exhibit maximum orthogonality (structural independence) \cite{ref25, ref26, ref28}.
\vspace{0.25cm}

\subsubsection{Sequential Least Squares Programming}
Capital allocation is treated as a non-linear convex optimisation problem solved using \textit{Sequential Least Squares Programming} (SLSQP) \cite{ref47, ref48, ref63}. The solver maximises the Sortino Ratio to target asymmetric upside capture.

\noindent \textbf{Downside Deviation:} Unlike standard deviation ($\sigma$), which penalises both upside and downside volatility, downside deviation ($DD$) measures risk only when returns fall below a target threshold (minimum acceptable return, $MAR$) \cite{ref29, ref30, ref31}. \textit{Annualised Downside Deviation} ($DD_p$) is defined as the root-mean-square of negative returns, scaled by the annual factor $A$ where $A=12$ for monthly optimisation)
\begin{equation}
DD_p(W) = \sqrt{A}\cdot\sqrt{\frac{1}{T}\sum_{t=1}^{T}\left[\min\left(0,r_{p,t}- \frac{R_f}{A}\right)\right]^2},
\end{equation}
where, $W$ is the vector of portfolio weights $[w_1,...,w_N]$, $r_{p,t}$ is the return of asset $i$ at time $t$.

\noindent \textbf{The Sortino Objective Function:} The optimisation objective function $F(w)$ is defined to maximise the excess return per unit of downside risk:
\begin{equation}
\text{Maximize } F(w) = \frac{1}{DD_p(w)}({w^T\mu-R_f}).
\end{equation}

\noindent \textbf{Constraints:} The optimisation is subject to strict boundary conditions to ensure feasibility and prevent concentration:
\begin{itemize}
    \item[(1)] Budget Constraint: $\sum_{i=1}^N w_i = 1$ (The portfolio is fully invested).
    \item[(2)] Long-Only Constraint: $0 \le w_i$ (Short selling is prohibited).
    \item[(3)] Diversification Cap: $w_i \le 0.05$ (Single asset exposure is capped at 5\%).
\end{itemize}

\subsection{Evaluation and Backtesting}
To rigorously assess the efficacy of the AEGIS model, a standardised backtesting engine was developed. This framework imposes strict constraints on transaction costs, rebalancing logic, and risk benchmarks to simulate realistic institutional trading conditions and prevent lookahead bias. \vspace{0.25cm}

\subsubsection{Backtesting Protocol}
The historical simulation operates on a walk-forward basis \cite{ref41, ref42, ref43}. For every rebalancing interval $t$, the model utilises a strictly retrospective training window $[t-L,t]$ to determine optimal weights, which are then applied to the unseen testing window $[t,t+1]$, where L is the number of look-back months.

\noindent \textbf{Rebalancing Frequency:} The portfolio is rebalanced according to the user-provided frequency. An optimal frequency balance (usually 1-3 months) is needed to respond to regime shifts (momentum decay), while minimising turnover costs.

\noindent \textbf{Transaction Cost Model (Friction):} To account for liquidity variance, bid-ask spreads, and commission fees, a proportional friction coefficient ($\delta$) is applied to the portfolio turnover.
\begin{equation}
NetReturn_t = GrossReturn_t - (\delta \times Turnover_t),
\end{equation}
where $\delta$ is usually set to 10 basis points (0.10\%), reflecting the cost of a trade in real life, and $Turnover_t$ is the sum of absolute weight changes $\sum |w_t - w_{t-1}|$.

\noindent \textbf{Risk-Free Rate ($R_f$):} A fixed annualised risk-free rate of 4.0\% is utilised for all risk-adjusted ratio calculations and as the hurdle rate for the \textit{sortino optimiser}. \vspace{0.25cm}

\subsubsection{Evaluation Metrics}
The strategy is evaluated using a suite of risk-adjusted metrics, calculated on the realised net returns of the testing window.

\noindent \textbf{Compound Annual Growth Rate:} The geometric progression ratio that provides a constant rate of return over the time period, smoothing out the volatility of periodic returns.
\begin{equation}
CAGR = \left(\frac{V_{final}}{V_{initial}}\right)^{\frac{1}{n}} - 1.
\end{equation}

\noindent \textbf{Sharpe Ratio:} Measures the average return earned in excess of the risk-free rate per unit of total risk (volatility).
\begin{equation}
Sharpe = \frac{1}{\sigma_p}({\bar{R}_p - R_f}),
\end{equation}
where $\bar{R}_p$ is the average annualised return and $\sigma_p$ is the annualised standard deviation of monthly returns.

\noindent \textbf{Sortino Ratio:} The primary objective function of the AEGIS optimiser. Unlike the sharpe ratio, the sortino ratio isolates downside volatility. It is calculated using the lower partial moment (LPM) of degree 2:
\begin{equation}
Sortino = \frac{1}{DD_{ann}}({R_{ann} - R_f}),
\end{equation}
where $R_{ann}$ is the annualised return and $DD_{ann}$ is the annualised downside deviation calculated on monthly testing data (say, $A = 12$):
\begin{equation}
DD_{ann} = \sqrt{12} \cdot \sqrt{\frac{1}{N} \sum_{i=1}^{N} \left[\min\left(0,r_i - \frac{R_f}{12}\right)\right]^2}.
\end{equation}

\noindent \textbf{Maximum Drawdown:} The maximum observed loss from a peak to a trough of the portfolio equity curve, before a new peak is attained. This serves as the primary stress test for tail-risk exposure \cite{ref32}.
\begin{equation}
MDD = \min_t \left( \frac{V_t - V_{peak}}{V_{peak}} \right).
\end{equation}

\noindent \textbf{Calmar Ratio:} A stringent measure of return relative to tail risk, calculated as the annualised return divided by the absolute value of the maximum drawdown. High calmar ratios indicate robust recovery characteristics \cite{ref33, ref34}.
\begin{equation}
Calmar = \frac{CAGR}{|MDD|}.
\end{equation}
\vspace{0.25cm}

\subsubsection{Bias Mitigation}
The model simulates an investor's experience in real-time by strictly segregating historical data into distinct Training (in-sample) and Testing (out-of-sample) windows. For each rebalancing interval $t$, the algorithm utilises a retrospective look-back period $[t - L,t]$ solely to generate signals and optimise portfolio weights. These fixed weights are then applied to the subsequent, unseen testing period $[t, t + 1]$ to calculate realised returns. Critically, the training window ends prior to the start of the testing window (e.g., training on Jan 1 to Dec 31, testing on Jan 1 to Feb 1), ensuring that the optimiser never accesses future price information. This temporal separation eliminates look-ahead bias \cite{ref44, ref46}, while applying a 10-basis-point friction model per transaction accounts for liquidity costs, providing a conservative estimate of net realisable performance.

\section{Proposed Model}

\subsection{Architecture Overview}
The AEGIS system architecture, illustrated in Figure 3, is designed as a modular, systematic equity selection engine.  To mitigate the structural weaknesses of traditional momentum strategies, specifically, their vulnerability to mean reversion and correlation breakdowns during market stress, the framework employs a strictly hierarchical, three-stage pipeline. This architecture enforces a separation of concerns between signal generation, risk management, and capital allocation.

\noindent \textbf{Signal Generation Module:} This module acts as the primary filter for the investment universe. It ingests raw price data and computes the VAM score for every asset in the S\&P Suite, NASDAQ-100, and Dow Jones. By normalising cumulative returns by their realised volatility, this layer isolates assets with genuine, idiosyncratic strength while rejecting high-beta stocks that are merely amplifying broad market movements.

\noindent \textbf{Immunisation Layer:} Operating downstream from the signal generator, this layer applies a minimax correlation filter. Its objective is not to maximise returns, but to minimise structural fragility. It iteratively constructs a portfolio that minimises the maximum pairwise correlation between any two assets. This mathematically forces the basket to be diversified across distinct risk factors, ensuring that the failure of one asset class (e.g., Tech) does not cascade through the entire portfolio.

\noindent \textbf{Allocation Engine:} The final stage is the rolling allocation engine, which determines the precise capital weight for each selected asset. Unlike mean-variance optimisers that penalise all volatility, this engine solves a non-linear convex optimisation problem to maximise the sortino ratio. This allows the portfolio to aggressively target upside volatility while strictly penalising downside deviation. The engine operates on a strict walk-forward basis, re-optimising weights monthly to adapt to changing market conditions without lookahead bias.

\begin{figure*}[htbp]
\centering
\includegraphics[width=\textwidth]{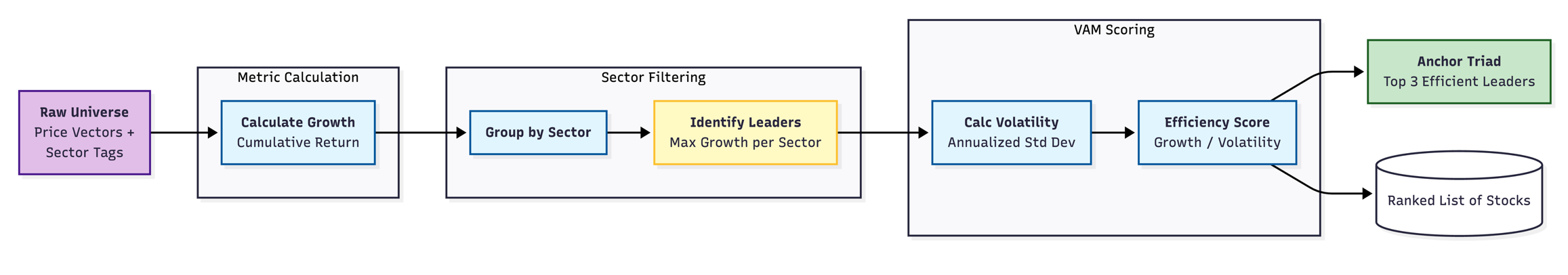}
\caption{The Signal Generation Module. This module executes a hierarchical selection protocol. First, the system ingests raw price data and identifies the Sector Leader for every GICS sector based purely on cumulative growth ($R_i$). These leaders are then subjected to VAM Scoring, where their returns are normalised by realised volatility to produce an Efficiency Score. The top three sector leaders with the highest Efficiency Scores are locked in as the Anchor Triad, ensuring the portfolio is constructed around the market's most robust structural trends.}
\label{fig:arch3a}
\end{figure*}
\begin{figure*}[htbp]
\centering
\includegraphics[width=\textwidth]{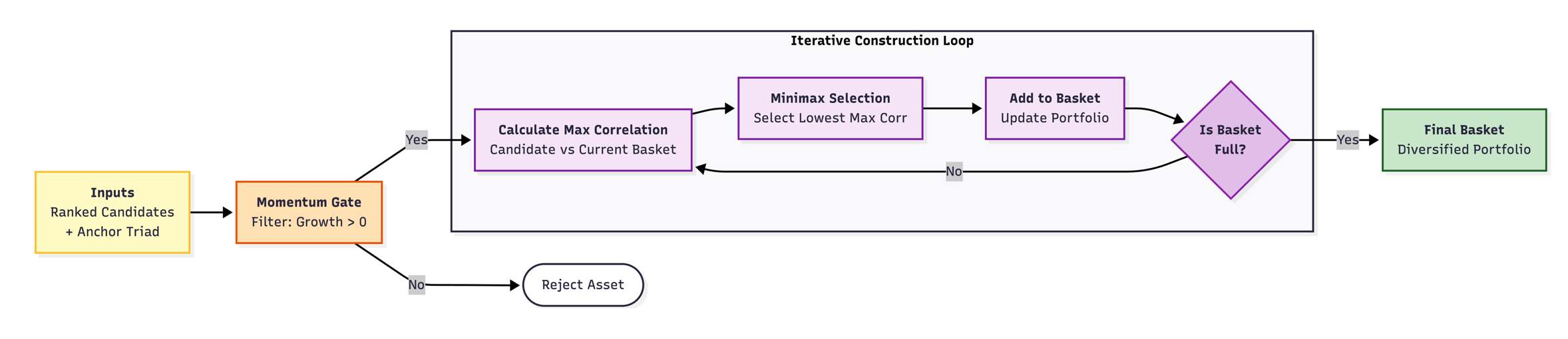}
\caption{The Immunisation Layer. This module constructs the portfolio iteratively to ensure robust diversification. First, the Momentum Gate filters the candidate pool, rejecting any asset with negative cumulative returns ($R_i<0$). The system then enters a Recursive Construction Loop: for each open slot, it calculates the maximum correlation of every remaining candidate against the current portfolio members. The candidate with the lowest maximum correlation is selected and added to the basket.}
\label{fig:archb}
\end{figure*}
\begin{figure*}[htbp]
\centering
\includegraphics[width=\textwidth]{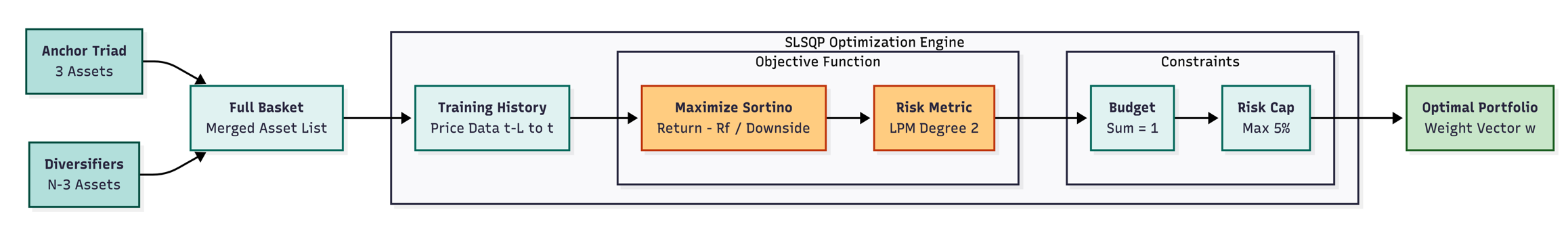}
\caption{The Allocation Engine. This module executes the final capital distribution. The Anchor Triad and selected Diversifiers are merged into a single Full Basket. The system then retrieves historical price data for the training window ($t-L \to t$) and feeds it into the SLSQP Optimisation Engine. The solver maximises the Sortino Ratio by minimising Downside Deviation (Lower Partial Moment of degree 2), subject to strict institutional constraints: a fully invested budget ($\sum w_i = 1$) and a maximum single-asset exposure of 5\% ($w_i \le 0.05$)}
\label{fig:archb}
\end{figure*}

\subsection{Signal Generation Module}
The primary objective of this phase is to identify the Efficiency Leaders, i.e., the assets that deliver superior risk-adjusted growth rather than raw price appreciation. This process, illustrated in Figure 3, executes a hierarchical filtration protocol. \vspace{0.25cm}

\subsubsection{Hierarchical Filtration Protocol}
The hierarchical filtration protocol is designed to choose the portfolio leading anchor stocks along with developing the ranked list of rest of the universe.

\noindent \textbf{Sector-Based Leader Identification:} To ensure the portfolio retains exposure to the market's primary structural drivers, the algorithm first groups the entire investment universe by GICS sector (e.g., Technology, Healthcare, Industrials). For each sector $S$, the asset with the highest cumulative return ($R_i$) over the look-back window $L$ (twelve months) is identified as the sector leader:
\begin{equation}
\text{Leader}_S = \arg\max_{i \in S}(R_{i,t-L:t}).
\end{equation}

\noindent \textbf{VAM Scoring:} Mere high returns are insufficient and the leaders must demonstrate stability. The algorithm computes the efficiency score ($S_{i,t}$) for each sector leader by normalising their cumulative return by their realised annualised volatility ($\sigma_i$):
\begin{equation}
S_{i,t} = \frac{R_{i,t-L:t-1}}{\sigma_i},
\end{equation}
where $\sigma_i = \sqrt{252} \cdot \text{std}(r_{i,t})$, is realised volatility over an entire year (assuming 252 working days).

\noindent \textbf{The Anchor Triad:} The top three sector leaders with the highest efficiency scores are locked into the portfolio as the anchor triad. These three assets form the offensive core of the basket, ensuring that the strategy is always aligned with the strongest trends in the broad market.

\begin{algorithm}
\caption{Hierarchical Filtration and Anchor Selection}\label{alg:1}
\begin{algorithmic}[1]
\State \textbf{Input:} Stock universe $U$, sector map $S$, look-back $L$
\State $Candidate\_List \gets []$
\State $Unique\_Sectors \gets \text{Unique}(S)$
\For{each sector $k$ in $Unique\_Sectors$}
    \State $U_k \gets \{asset\ i \mid S(i)=k\}$
    \State $Leader_k \gets \arg\max(\text{CumReturn}(i, L)) \forall i \in U_k$
    \State $R_k \gets \text{CumReturn}(Leader_k, L)$
    \State $\Sigma_k \gets \text{AnnVol}(Leader_k, L)$
    \State $Efficiency_k \gets R_k/\Sigma_k$
    \State Add $(Leader_k, Efficiency_k)$ to $Candidate\_List$
\EndFor
\State Sort $Candidate\_List$ descending by $Efficiency_k$
\State $A \gets \text{Top 3 Assets from } Candidate\_List$
\State \textbf{Return} $A$
\end{algorithmic}
\end{algorithm}

\subsection{The Immunisation Layer}
Once the anchor triad is established, the remaining capital slots (the diversifiers) must be filled. The objective here is strictly risk minimisation. This module, shown in Figure 4, constructs a basket that is mathematically orthogonal to the anchors.  \vspace{0.25cm}

\subsubsection{Structural Diversification Protocol}
This protocol ensures the the portfolio is well diversified and ensures the portfolio is protected against market crashes.

\noindent \textbf{The Momentum Gate:} Before any correlation analysis, the candidate pool is subjected to a boolean momentum gate. An asset is eligible for selection only if its cumulative return is positive ($R_i > 0$). This crucial step prevents the value trap that is common in low-correlation strategies, where the algorithm might otherwise select distressed assets simply because they are uncorrelated with the market.

\noindent \textbf{Recursive Minimax Filtration:} The system employs a greedy iterative algorithm to fill the remaining slots ($N-3$). 

\noindent Initialisation: The basket $B$ starts with the anchor triad.

\noindent Iteration: For every remaining candidate $c$ in the eligible pool, the algorithm calculates its maximum pairwise correlation against all current members of basket $B$:
\begin{equation}
\rho_{\max}(c) = \max_{b \in B} |\text{corr}(r_c, r_b)|.
\end{equation}

\noindent Selection: The candidate with the lowest $\rho_{\max}$ is added to basket $B$:
\begin{equation}
c^* = \arg\min_c (\rho_{\max}(c)).
\end{equation}

\begin{algorithm}
\caption{Structural Diversification Protocol}\label{alg:2}
\begin{algorithmic}[1]
\State \textbf{Input:} Anchor triad $A$, candidate pool $U$, target size $N$
\State $B \gets A$
\State $P \gets \{\}$
\For{each asset $i$ in $U$}
    \If{$i \notin A$ \textbf{and} $\text{CumReturn}(i) > 0$}
        \State Add $i$ to $P$
    \EndIf
\EndFor
\While{$\text{Size}(B) < N$}
    \State $Best\_Asset \gets \text{Null}$
    \State $Min\_Max\_Corr \gets \infty$
    \For{each candidate $c$ in $P$}
        \State $Rho\_max \gets \max(|\text{Corr}(c, b)|) \forall b \in B$
        \If{$Rho\_max < Min\_Max\_Corr$}
            \State $Min\_Max\_Corr \gets Rho\_max$
            \State $Best\_Asset \gets c$
        \EndIf
    \EndFor
    \If{$Best\_Asset \neq \text{Null}$}
        \State Add $Best\_Asset$ to $B$
        \State Remove $Best\_Asset$ from $P$
    \Else
        \State \textbf{Break}
    \EndIf
\EndWhile
\State \textbf{Return} $B$
\end{algorithmic}
\end{algorithm}

\noindent Recursion: The matrix is updated, and the process repeats until the basket reaches the target size (e.g., $N=50$). This recursive logic ensures that every new addition is specifically chosen to dampen the systemic risk of the current portfolio. The immunisation layer functions as the architectural safeguard of the AEGIS framework, transforming the concentrated beta of the anchor triad into a structurally resilient, anti-fragile portfolio. By mathematically enforcing orthogonality across the final constituent basket, the system guarantees that systemic shocks affecting a single sector or risk factor are locally contained rather than cascaded. This proactive decoupling of correlated risks neutralises the \textit{``Winner's Curse,''} establishing a highly stable, low-covariance foundation optimally primed for the subsequent convex allocation phase.

\begin{algorithm}
\caption{Weight Allocation via Convex Optimisation}\label{alg:3}
\begin{algorithmic}[1]
\State \textbf{Input:} Basket $B$, history $H$, look-back $L$, friction $f$
\State $t \gets T_0$
\State $w_{prev} \gets [0, 0, \dots, 0]$
\While{$t < EndOfData$}
    \State $Train\_Prices \gets H[t-L:t]$
    \State $Test \gets H[t:t+1]$
    \State $R \gets (Train\_Prices[1:] / Train\_Prices[:-1]) - 1$
    \Function{Objective}{$w$}
        \State $Daily\_Port\_Ret \gets R \times w$
        \State $Ann\_Ret \gets \text{Mean}(Daily\_Port\_Ret) \times 252$
        \State $Downside \gets \min(Daily\_Port\_Ret - (0.04/252), 0)$
        \State $LPM\_2 \gets \sqrt{\text{Mean}(Downside^2)} \times \sqrt{252}$
        \State \textbf{Return} $-(Ann\_Ret - 0.04) / (LPM\_2 + \epsilon)$
    \EndFunction
    \State $Constraints \gets [\sum w = 1, 0 \le w \le 0.05]$
    \State $w_{opt} \gets \text{Minimise}(\text{Objective}, Constraints)$
    \State $t\_diff \gets (Test[End] - Test[Start]) / Test[Start]$
    \State $Period\_Ret\_Asset \gets t\_diff$
    \State $Gross\_Ret \gets \sum (w_{opt} \times Period\_Ret\_Asset)$
    \State $Turnover \gets \sum |w_{opt} - w_{prev}|$
    \State $Cost \gets Turnover \times f$
    \State Store $(Gross\_Ret - Cost)$
    \State $w_{prev} \gets w_{opt}$
    \State $t \gets t + 1 \text{ Month}$
\EndWhile
\State \textbf{Return} Series of $R\_net$
\end{algorithmic}
\end{algorithm}

\subsection{The Allocation Engine}
The final phase, detailed in Figure 5, determines the precise capital weight ($w_i$) for each asset in the constructed basket.  Unlike heuristic weighting schemes (e.g., 1/N or risk parity), AEGIS employs a non-linear convex optimisation engine \cite{ref63}. \vspace{0.25cm}

\subsubsection{Convex Optimisation Protocol}
The convex optimisation protocol is responsible for the final optimal allocation of the assets based on the look-back duration provided by the user.

\noindent \textbf{The Optimisation Window:} To strictly eliminate look-ahead bias, the optimiser operates on a retrospective training window $[t-L, t]$. The weights derived from this window are then frozen and applied to the testing window $[t, t+1]$ to realise returns.

\noindent \textbf{Objective Function:} The solver is configured to maximise the sortino ratio, effectively treating upside volatility as beneficial while penalising downside deviation.
\begin{equation}
\text{Maximize } F(w) = \frac{w^T\mu - R_f}{DD_{ann}},
\end{equation}
where $DD_{ann}$ is the annualised lower partial moment (LPM) of degree 2, calculated using a risk-free hurdle rate of 4.0\% \cite{ref34}.

\begin{table*}[tbp]
\centering
\caption{Detailed annual performance and risk profile of the AEGIS strategy from 2006 to 2025. The data illustrates the consistent constraint of transaction friction (averaging $<1.0\%$ annually) alongside realised net returns. Key regime-adaptive behaviours are evident in the contrast between high-momentum bull markets (e.g., 2013: +51.24\% net and negligible drawdown) and systemic shock environments (e.g., 2008: limiting drawdowns to -28.00\% during the Global Financial Crisis).}
\label{tab:annual_perf}
\resizebox{\textwidth}{!}{%
\begin{tabular}{@{}cccccccccc@{}}
\toprule
\textbf{Year} & \textbf{Basket Size} & \textbf{Gross Return} & \textbf{Friction} & \textbf{Net Return} & \textbf{Avg Monthly} & \textbf{Annual Vol} & \textbf{Sortino} & \textbf{Win Rate} & \textbf{Max DD} \\ \midrule
2006 & 50 & 20.23\% & -0.82\% & 19.26\% & 1.51\% & 9.08\% & 4.28 & 66.7\% & -3.51\% \\
2007 & 50 & 16.96\% & -0.77\% & 16.07\% & 1.31\% & 12.38\% & 1.46 & 75.0\% & -7.23\% \\
2008 & 50 & -20.29\% & -0.80\% & -20.94\% & -1.79\% & 18.86\% & -1.45 & 33.3\% & -28.00\% \\
2009 & 50 & 3.53\% & -0.75\% & 2.76\% & 0.34\% & 16.48\% & 0.01 & 66.7\% & -12.08\% \\
2010 & 50 & 27.25\% & -0.74\% & 26.32\% & 2.06\% & 15.40\% & 2.61 & 75.0\% & -8.89\% \\
2011 & 50 & 3.91\% & -0.80\% & 3.09\% & 0.33\% & 13.96\% & 0.00 & 58.3\% & -13.76\% \\
2012 & 50 & 33.55\% & -0.77\% & 32.55\% & 2.45\% & 13.41\% & 5.50 & 75.0\% & -3.65\% \\
2013 & 50 & 52.45\% & -0.83\% & 51.24\% & 3.53\% & 7.46\% & 82.61 & 91.7\% & -0.13\% \\
2014 & 50 & 16.94\% & -0.78\% & 16.05\% & 1.31\% & 12.63\% & 1.76 & 66.7\% & -4.68\% \\
2015 & 50 & 10.48\% & -0.77\% & 9.64\% & 0.83\% & 12.48\% & 0.74 & 66.7\% & -5.82\% \\
2016 & 50 & 14.31\% & -0.80\% & 13.40\% & 1.17\% & 16.97\% & 0.89 & 66.7\% & -10.40\% \\
2017 & 50 & 22.71\% & -0.82\% & 21.72\% & 1.67\% & 5.85\% & 15.79 & 91.7\% & -0.61\% \\
2018 & 50 & 1.29\% & -0.85\% & 0.42\% & 0.14\% & 15.85\% & -0.18 & 58.3\% & -16.73\% \\
2019 & 50 & 21.56\% & -0.72\% & 20.71\% & 1.61\% & 8.77\% & 5.49 & 75.0\% & -1.96\% \\
2020 & 50 & 54.16\% & -0.72\% & 53.06\% & 4.23\% & 38.30\% & 2.00 & 75.0\% & -28.89\% \\
2021 & 50 & 18.68\% & -0.85\% & 17.69\% & 1.42\% & 10.94\% & 2.14 & 75.0\% & -3.78\% \\
2022 & 50 & -4.38\% & -0.74\% & -5.10\% & -0.23\% & 22.06\% & -0.41 & 50.0\% & -13.65\% \\
2023 & 50 & 15.04\% & -0.74\% & 14.20\% & 1.15\% & 8.90\% & 2.01 & 58.3\% & -5.85\% \\
2024 & 50 & 25.26\% & -0.71\% & 24.40\% & 1.93\% & 15.01\% & 1.97 & 83.3\% & -6.84\% \\
2025 & 50 & 16.69\% & -0.80\% & 15.77\% & 1.29\% & 12.41\% & 2.13 & 66.7\% & -6.36\% \\ \bottomrule
\end{tabular}%
}
\end{table*}

\noindent \textbf{Institutional Constraints:} To ensure the resulting portfolio is tradable and compliant with standard risk mandates, the optimisation is bound by three strict constraints:
\begin{itemize}
    \item Budget Constraint: $\sum w_i = 1$ (the portfolio must be fully invested).
    \item Long-Only Constraint: $w_i \ge 0$ (short selling is prohibited).
    \item Diversification Cap: $w_i \le 0.05$ (no single asset may exceed 5\% of total equity). 
\end{itemize}
This constrained optimisation forces the engine to spread capital efficiently across the most stable assets in the basket, rather than concentrating risk in a single winner. These caps help provide the structural soundness of the framework.

\subsection{Theoretical Conclusion}
The AEGIS framework, by integrating algorithms 1, 2, and 3, establishes a rigorous quantitative pipeline that addresses the structural limitations of traditional momentum strategies. By coupling the VAM signal (which isolates idiosyncratic alpha) with the minimax correlation filter (which enforces orthogonality), the architecture mathematically prevents the risk-clustering phenomenon often observed during market corrections. Furthermore, the application of the sortino-based allocation engine ensures that the portfolio remains aggressive in capturing upside volatility while strictly penalising downside deviation. The following section, empirical results, validates the efficacy of this architecture through a multi-year walk-forward simulation across the U.S. equity universe.

\section{Results and Discussions}

\subsection{Annualised Consistency}
Before evaluating the long-term capital compounding against the broad market benchmarks, it is critical to analyse the AEGIS framework’s behaviour on a granular, year-by-year basis. Table II details the annual gross returns, transaction friction, realised net performance, and underlying risk metrics of the strategy across the 20-year testing horizon. Here, we used a 3-month look-back for allocation, along with 47 diversifiers. \vspace{0.25cm}

\subsubsection{Optimiser Stability and Low Friction}
A persistent failure point for systematic momentum strategies is excessive portfolio turnover, which erodes the alpha generated. The SLSQP allocation engine demonstrates extreme stability. Over two decades of rolling monthly rebalances, the total accumulated friction was constrained to just -15.58\%. This operational efficiency allowed the model to convert a massive 1950.78\% gross return into a highly realisable 1657.17\% net return, confirming that the optimiser successfully holds long-term winners and avoids algorithmic over-trading. \vspace{0.25cm}

\begin{figure*}[tbp]
\centering
    \includegraphics[width=\textwidth]{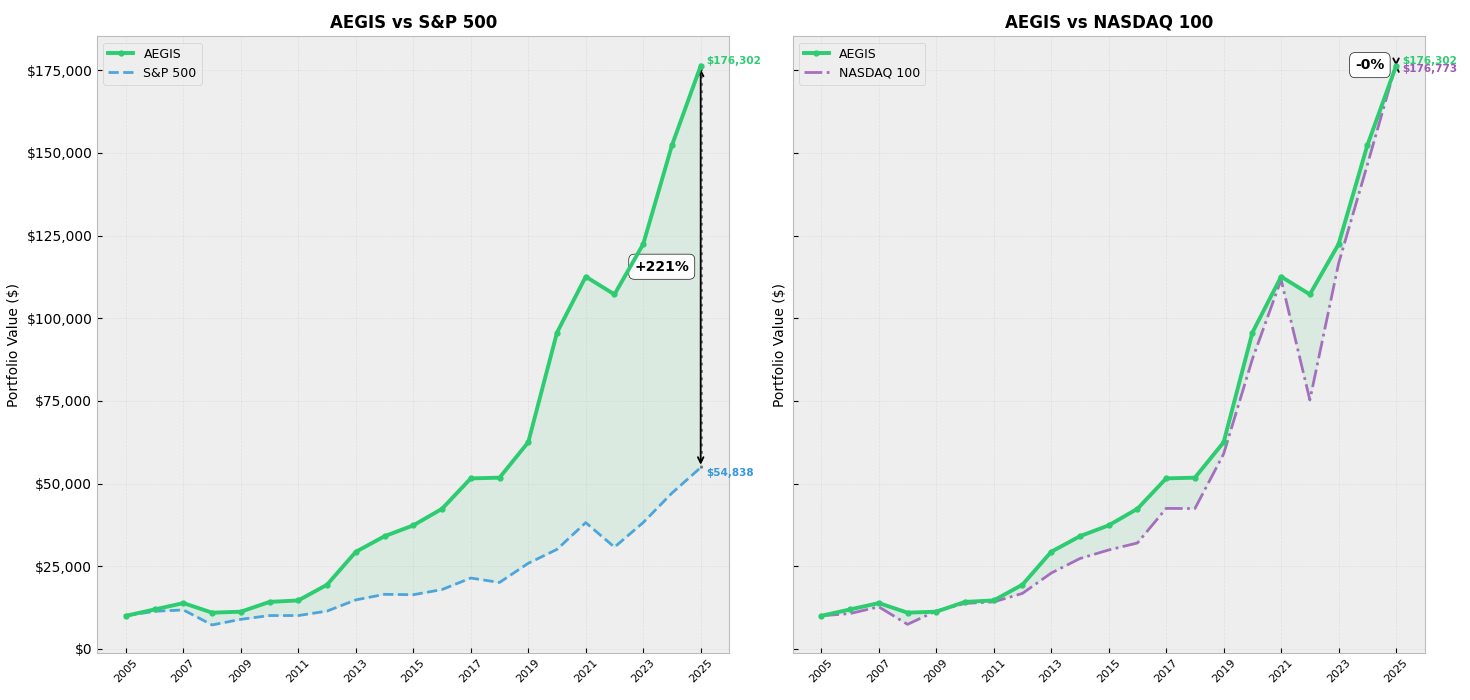}
    \caption{Graph 1 and 2: 20-Year Capital Appreciation: AEGIS Broad Market Dominance (S\&P 500) and Tech-Parity (NASDAQ) Trajectory (\$10,000 Initial Allocation)}

\label{fig:graphs1_2}
\end{figure*}

\begin{table}[htbp]
\centering
\caption{Overall Statistics from a 20-year walk-forward backtest from 2006 to 2025 covering multiple regimes and major financial events.}
\label{tab:overall_stats}
\begin{tabular}{@{}ll@{}}
\toprule
\textbf{Category} & \textbf{Metric / Value} \\ \midrule
\multirow{4}{*}{\textbf{Returns}} & Total Gross Return: 1950.78\% \\
 & Total Net Return: 1657.17\% \\
 & CAGR: 15.41\% \\
 & Total Friction Impact: -15.58\% \\ \midrule
\multirow{5}{*}{\textbf{Risk}} & Annualized Volatility: 16.44\% \\
 & Maximum Drawdown: -28.89\% \\
 & Sharpe Ratio: 0.72 \\
 & Average Sortino Ratio: 6.47 \\
 & Outlier Adjusted Sortino: 1.72 \\ \midrule
\multirow{5}{*}{\textbf{Consistency}} & Monthly Win Rate: 68.8\% \\
 & Profitable Years: 18/20 \\
 & Best Year (2020): 53.06\% \\
 & Worst Year (2008): -20.94\% \\
 & Average Monthly Gain: 1.31\% \\ \bottomrule
\end{tabular}
\end{table}

\subsubsection{Win Rate and Baseline Consistency}
The framework exhibits remarkable consistency, achieving profitability in 90\% of the tested years (18 out of 20). By generating a positive return in 68.8\% of all traded months, the strategy produced a steady average monthly return of 1.31\%. This high frequency of localised success indicates that the VAM filter effectively eliminates erratic assets, yielding a mathematically smooth equity curve. \vspace{0.25cm}

\subsubsection{Asymmetric Risk Profile}
The ultimate objective of the AEGIS framework is to engineer an asymmetric risk-to-reward profile, capturing upside velocity while rejecting downside variance. The aggregate risk metrics explicitly validate this: an annualised volatility of 16.44\% coupled with a heavily constrained maximum drawdown of -28.89\% yields a robust sharpe ratio of 0.72. However, the true efficacy of the downside penalisation is captured by the exceptional average sortino ratio of 6.47. However, the outlier adjust sortino ratio for the same stands at 1.72 over the 18 years (two outliers i.e., years with exceptional performance). This asymmetry is vividly illustrated by contrasting extreme market regimes. During systemic expansions (e.g., 2020), the strategy achieved maximum upside capture (+53.06\%). Conversely, during the 2008 global financial crisis, the minimax immunisation layer acted as a defensive circuit breaker, restricting the model's worst historical year to -20.94\% and successfully insulating the portfolio against total market collapse.

\begin{figure*}[tbp]
\centering
    \includegraphics[width=\textwidth]{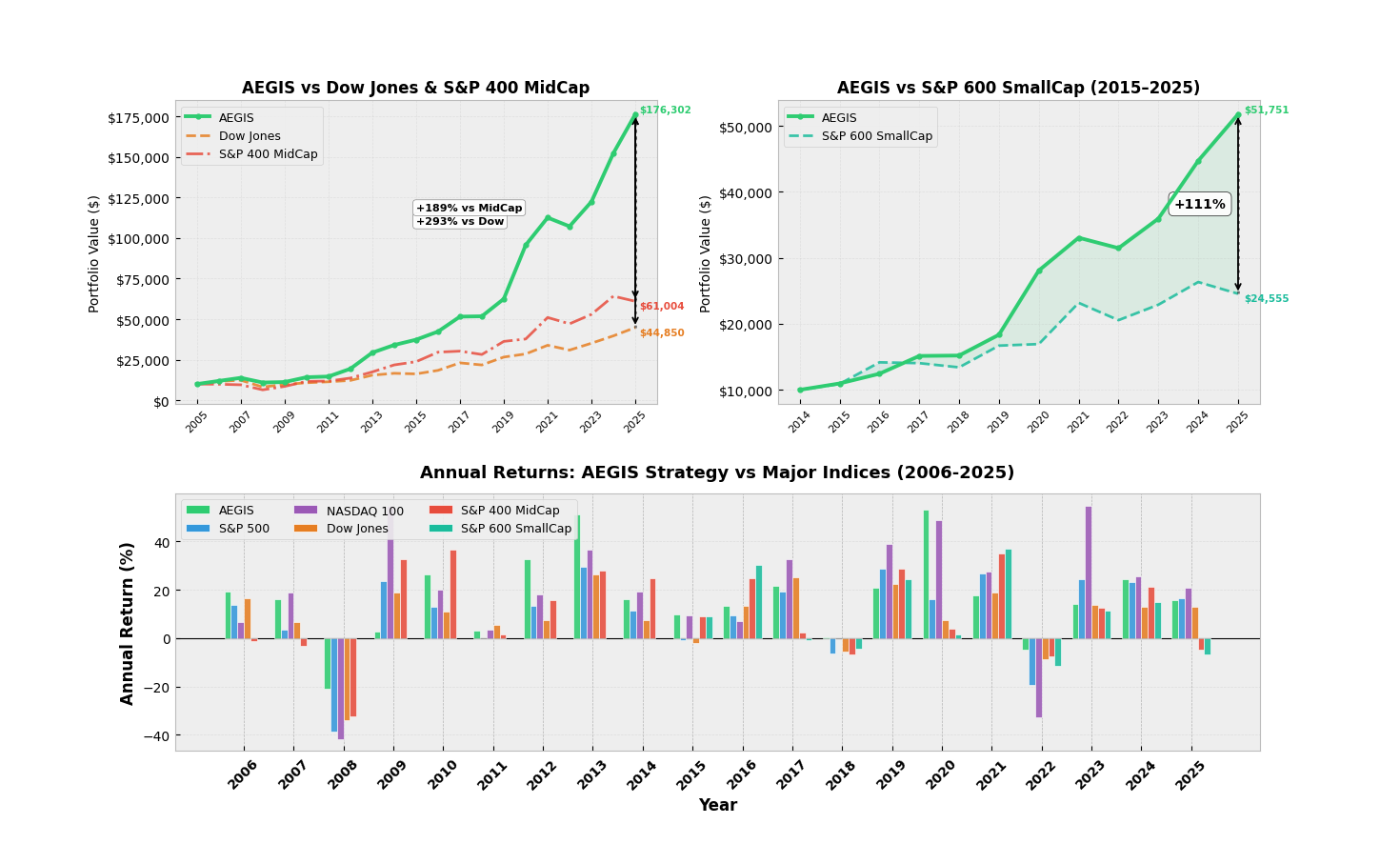}
    \caption{Graph 3 and 4: 20-Year Capital Appreciation: AEGIS vs. Dow Jones, S\&P 400 MidCap, and S\&P 600 SmallCap Indices. This represents Terminal Wealth Divergence and Size-Factor Independence as per general conception. Graph 5: Overall visualisation of gain/loss generated by AEGIS versus all the major U.S. Benchmarks over the entire 20-year backtest period. Note: The S\&P 600 SmallCap index data has been added since 2015 due to lack of pre-2015 data of returns}

\label{fig:graphs3_4_5}
\end{figure*}

\begin{table*}[tbp]
\centering
\caption{Multi-Horizon Compound Annual Growth Rate (CAGR) Comparison of AEGIS Versus Major US. Equity Benchmarks (Ending 2025).}
\label{tab:multi_horizon}
\begin{tabular}{@{}llcccccc@{}}
\toprule
\textbf{Time Horizon} & \textbf{Time Range} & \textbf{AEGIS Strategy} & \textbf{S\&P 500} & \textbf{NASDAQ-100} & \textbf{Dow Jones} & \textbf{S\&P 400 MidCap} & \textbf{S\&P 600 SmallCap} \\ \midrule
1-Year & 2025-2025 & 15.77\% & 16.39\% & 20.77\% & 12.97\% & 4.95\% & -6.74\% \\
2-Year & 2024-2025 & 20.01\% & 19.80\% & 23.15\% & 12.92\% & 7.32\% & 3.57\% \\
5-Year & 2021-2025 & 12.93\% & 12.76\% & 15.07\% & 9.45\% & 10.07\% & 7.72\% \\
10-Year & 2016-2025 & 16.75\% & 12.85\% & 19.45\% & 10.68\% & 9.88\% & 8.47\% \\
15-Year & 2011-2025 & 18.26\% & 11.96\% & 18.57\% & 9.96\% & 11.66\% & - \\
20-Year & 2006-2025 & 15.41\% & 8.88\% & 15.44\% & 7.79\% & 9.46\% & - \\ \bottomrule
\end{tabular}
\end{table*}

\subsection{Multi-Horizon Benchmarking}
Having established the internal stability and risk-adjusted efficiency of the AEGIS architecture, it is essential to contextualise this performance against traditional passive investment vehicles. To test for regime dependency, the risk that a strategy's success is merely an artefact of a specific, localised bull market, the framework's CAGR was evaluated across rolling 1, 2, 5, 10, 15, and 20-year horizons ending in 2025. The strategy was benchmarked against the broad-market \textit{S\&P} 500, the technology-concentrated \textit{NASDAQ-100}, the blue-chip \textit{Dow Jones Industrial Average}, the \textit{S\&P 400 MidCap} and the \textit{S\&P 600 SmallCap} indices. Over the complete 20-year lifecycle, AEGIS compounded at 15.41\% annually, nearly doubling the annualised return of the standard \textit{S\&P 500} (8.88\%) and the \textit{Dow Jones} (7.79\%). This massive divergence in terminal wealth confirms the success of the signal generation module in identifying high-efficiency assets while filtering out the structural drag of underperforming market sectors.

While the technology-heavy \textit{NASDAQ-100} posted formidable absolute returns, particularly in shorter 1 to 2 year horizons driven by extreme sector concentration, this precise lack of diversification exposed the benchmark to catastrophic vulnerability during systemic shocks. Severe tech-led drawdowns, most notably the -41.73\% collapse during the 2008 global financial crisis and the -32.58\% routing during the 2022 macroeconomic tightening cycle, created a massive volatility drag that severely degraded the index's long-term compounding efficiency. By successfully truncating these extreme capital destructions through its immunisation layer, AEGIS achieved virtual return parity with the index over the full two decades (15.41\% vs. 15.44\% for NASDAQ). Crucially, the model accomplished this without assuming the binary concentration risk inherent to the NASDAQ, utilising a dynamically diversified basket to engineer tech-like growth alongside utility-like downside protection.

In quantitative finance, outperformance is often scrutinised as a byproduct of assuming excessive risk in lower-tier market capitalisations \cite{ref11, ref12, ref13}. The inclusion of the MidCap and SmallCap indices definitively disproves this for AEGIS. Over the 10-year horizon, AEGIS (16.75\%) dramatically outpaced both the MidCap (9.88\%) and SmallCap (8.47\%) indices. Furthermore, during the recent 1-year horizon (2025), where the broader MidCap (-4.95\%) and SmallCap (-6.74\%) indices suffered structural declines, the AEGIS VAM filter successfully rotated capital away from decaying small-cap assets, preserving a +15.77\% gain.

\subsection{Cumulative Wealth Accumulation and Trajectory Analysis}
While annualised compounding metrics provide a standardised measure of regime independence, evaluating the absolute capital appreciation trajectory is necessary to understand the structural advantages of the AEGIS equity curve. To visualise this compounding efficiency, a baseline \$10,000 initial investment was simulated across the strategy and the benchmark indices over the specified horizons, assuming frictionless reinvestment of all generated alpha. \vspace{0.25cm}

\begin{figure*}[tbp]
\centering
    \includegraphics[width=\textwidth]{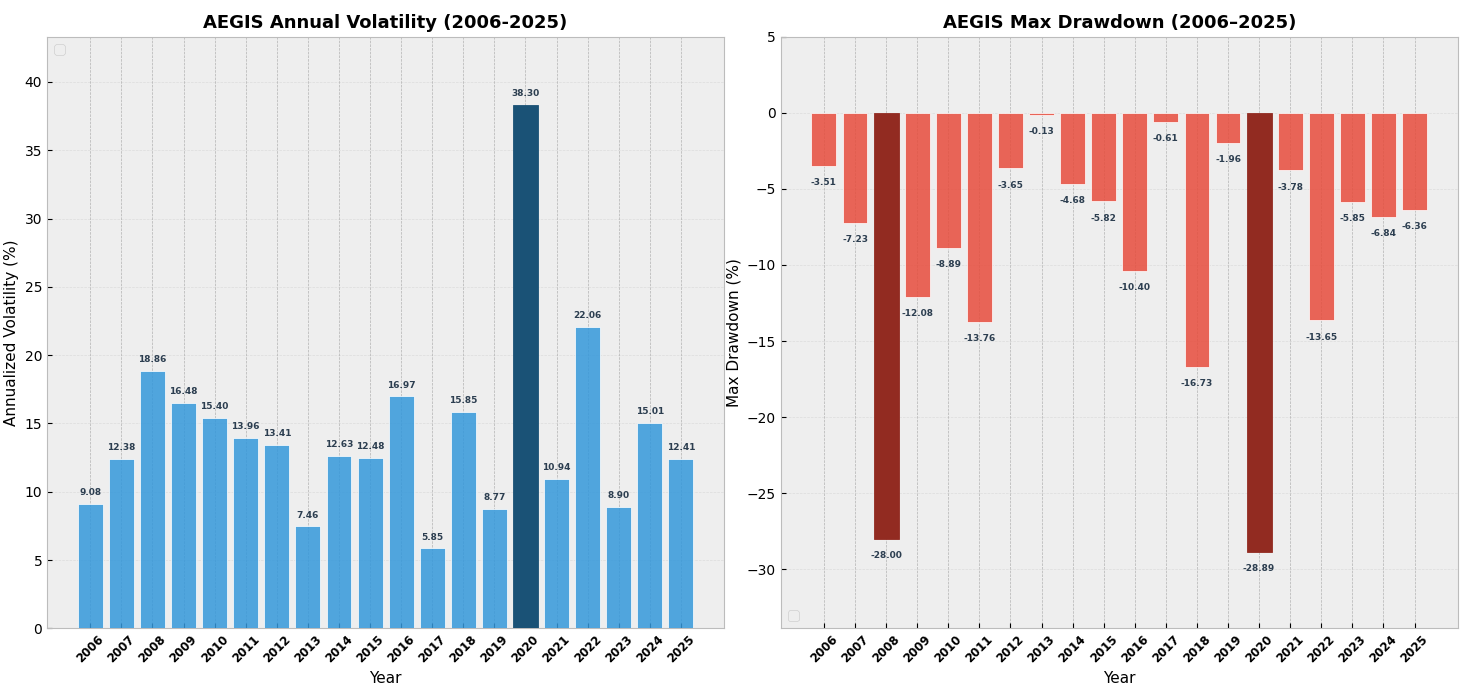}
    \caption{Graph 6 and 7: Year-wise annual volatility and max drawdown of AEGIS framework during the 20-year walk-forward backtest with the marked average annualised volatility and the average max drawdown over 20-years}
\label{fig:graphs6_7}
\end{figure*}

\subsubsection{Exponential Broad Market Alpha}
The comparison against the S\&P 500 in Graph 1 illustrates the sheer magnitude of wealth divergence generated by the allocation engine. Over the two-decade simulation, the standard market benchmark grew the initial \$10,000 allocation to \$54,838. In stark contrast, the AEGIS framework aggressively compounded the identical starting capital to \$176,302. The expanding shaded region highlights a massive +221\% terminal outperformance, graphically proving the long-term impact of consistently rotating into high-efficiency momentum assets while avoiding broad-market structural drag. \vspace{0.25cm}

\subsubsection{Risk-Adjusted Tech Parity}
The comparison against the \textit{NASDAQ-100} in Graph 2 offers the most profound visual evidence of the framework’s risk-adjusted superiority. While both the NASDAQ and AEGIS converge at a virtually identical terminal wealth (\$176,000), their respective pathways are fundamentally different. The NASDAQ's trajectory (the dashed purple line) is characterised by violent parabolic expansions immediately followed by severe, wealth-destroying corrections, most visibly during the catastrophic 2022 macroeconomic tech drawdown. Conversely, the AEGIS equity curve (the solid green line) traces a remarkably smoother, more monotonic climb. The visual confirms that AEGIS successfully matches the hyper-growth of the world’s most aggressive technology index, but achieves it through a dynamically hedged basket that avoids the NASDAQ's binary concentration risk and extreme volatility drag. \vspace{0.25cm}

\subsubsection{Rejection of the Size Premium}
The comparison against the \textit{S\&P 600 SmallCap} index in Graph 4 over the trailing 11-year window provides the ultimate visual refutation of the Fama-French size factor argument \cite{ref11}. If the AEGIS model's high returns were simply a proxy for taking on risky, small-capitalisation exposure, its equity curve would tightly correlate with the S\&P 600. Instead, the chart reveals a profound structural decoupling. From 2021 to 2025, the SmallCap index experienced a prolonged period of stagnation and decay, ending the period at \$24,555. During this exact same macroeconomic window, the AEGIS actively rotated capital away from these deteriorating small-cap assets, pushing its terminal wealth to \$51,751, a staggering +111\% outperformance. 

Additionally over the full lifecycle, the traditional blue-chip Dow Jones index suffered from severe structural lag, yielding a terminal value of only \$44,850 as seen in Graph 3. While the MidCap index initially demonstrated competitive growth, it experienced significant volatility drag post-2021, capping its terminal wealth at \$61,004. AEGIS dynamically bypassed the stagnation of these specific sectors, generating a massive +293\% outperformance over the Dow and a +189\% outperformance over the MidCap index.

\subsection{Risk Management, Drawdown Analysis, and Crisis Resilience}
While capital appreciation confirms the proper working of the VAM signal, institutional viability is ultimately determined by capital preservation during severe market dislocations. To evaluate the strategy’s asymmetric risk profile, the rolling annualised volatility and maximum drawdown were tracked across the 240-month testing horizon as plotted in graphs 6 and 7 respectively. The synthesis of the strategy's variance metrics reveals a profound structural advantage generated by the minimax immunisation layer, specifically in its ability to truncate left-tail risk (systemic losses) while allowing right-tail compounding (momentum profits) to run unhindered.
\vspace{0.25cm}

\begin{figure*}[tbp]
\centering
    \includegraphics[width=\textwidth]{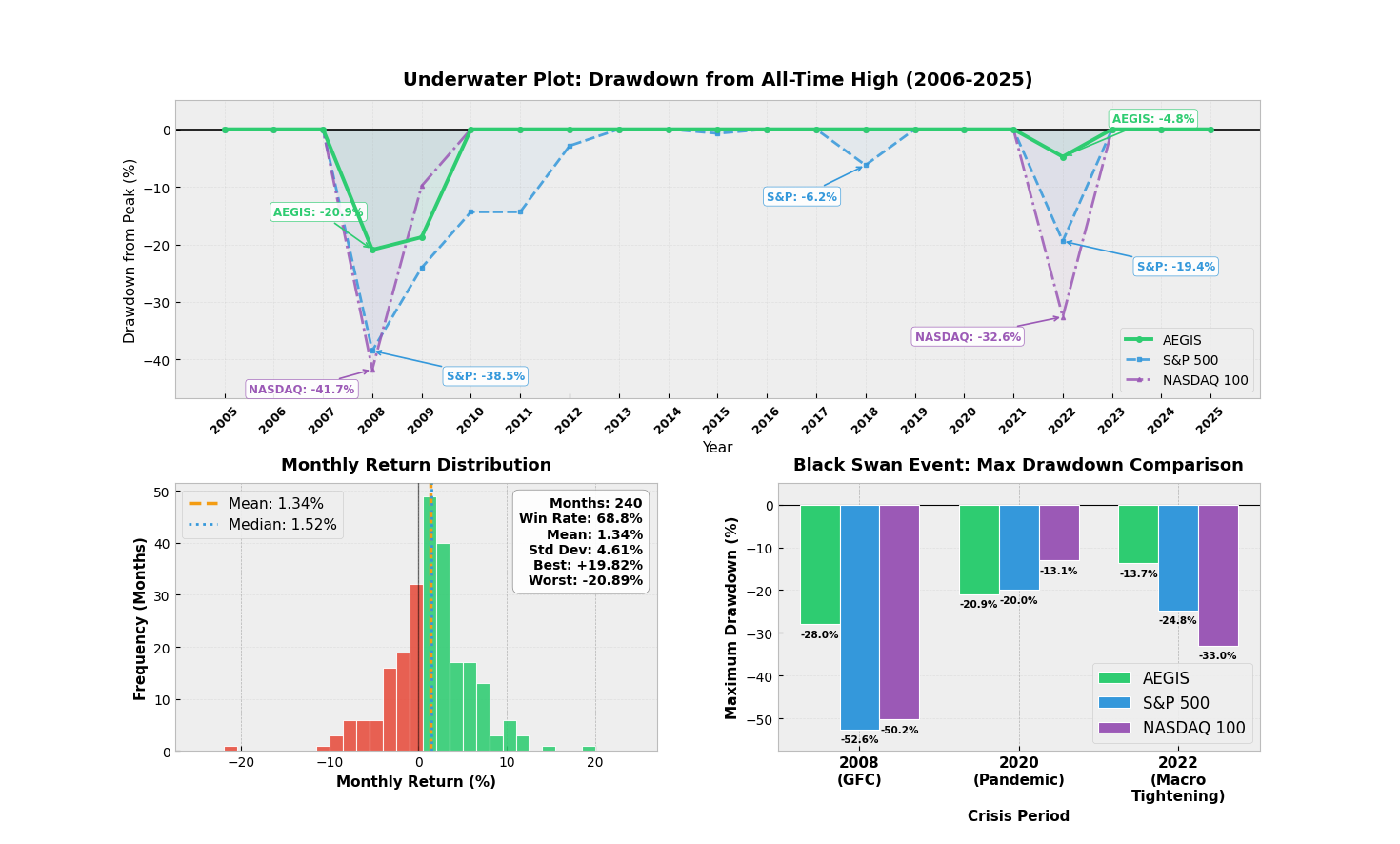}
    \caption{Graph 8: Comparative Underwater Drawdown Analysis (2006-2025). The plot tracks the percentage decline from historical high-water marks, illustrating the framework's superior capital preservation and shallower recovery depths during the 2008 Global Financial Crisis and the 2022 macroeconomic tightening cycle. Graph 9 and 10: Monthly Return Distribution Skewness over the 20-year walk forward backtest and the Maximum Drawdown comparison across Major Systemic Shocks (2006–2025) e.g., the Global Financial Crisis (GFC) in 2008, with two of the major indices (S\&P 500 and NASDAQ).}
\label{fig:graphs8_9_10}
\end{figure*}

\subsubsection{Outlier Adjusted Sortino Profile}
Standard risk metrics, such as the sharpe ratio (0.72), penalise all volatility equally. However, because the AEGIS architecture is designed to capture explosive upside momentum, upside volatility is a desired feature, not a flaw. The sortino ratio, which exclusively penalises downside variance, provides a much more accurate representation of the framework’s risk efficiency. Over the 20-year horizon, AEGIS achieved a staggering average sortino ratio of 6.47. However, a rigorous statistical review requires addressing extreme positive outliers within the dataset. In 2013 and 2017, the model locked onto perfectly monotonic bullish trends, generating substantial net returns (+51.24\% and +21.72\%, respectively) while experiencing near-zero intra-year drawdowns (-0.13\% and -0.61\%). Because the sortino formula divides returns by downside deviation, these mathematically negligible drawdowns produced anomalous sortino spikes of 82.61 (2013) and 15.79 (2017). To present the most conservative baseline for institutional evaluation, these two extreme outliers were removed. The resulting outlier-adjusted average sortino ratio is 1.72. In quantitative finance, a sustained baseline sortino above 1.0 is considered excellent; a baseline of 1.72 over 18 years confirms that the algorithm reliably generates persistent alpha with tightly constrained downside variance, entirely independent of its most perfect years. \vspace{0.25cm}

\subsubsection{Tail-Risk Truncation and Return Asymmetry Analysis}
To visually validate the mathematical findings of the sortino analysis, the frequency distribution of the strategy's monthly returns was mapped over the complete 240-month lifecycle. In a standard, passively managed equity portfolio, return distributions typically resemble a normal bell curve with fat left tails, indicating a high susceptibility to extreme market drawdowns. As evidenced in Graph 9, the AEGIS distribution exhibits a profound right-skew, graphically confirming the framework's positive expectancy. The algorithm generated a positive return in 68.8\% of all traded months, yielding a median monthly return of +1.52\%. More critically, the tight clustering of negative returns near the zero-bound proves that the minimax immunisation layer successfully executes its primary directive i.e., aggressively cutting localised losses before they cascade into structural drawdowns. By truncating the left tail of the distribution, the architecture allows upside momentum to compound unhindered, successfully bridging the gap between theoretical asymmetry and live-market execution \cite{ref62}. \vspace{0.25cm}

\begin{figure*}[tbp]
\centering
    \includegraphics[width=\textwidth]{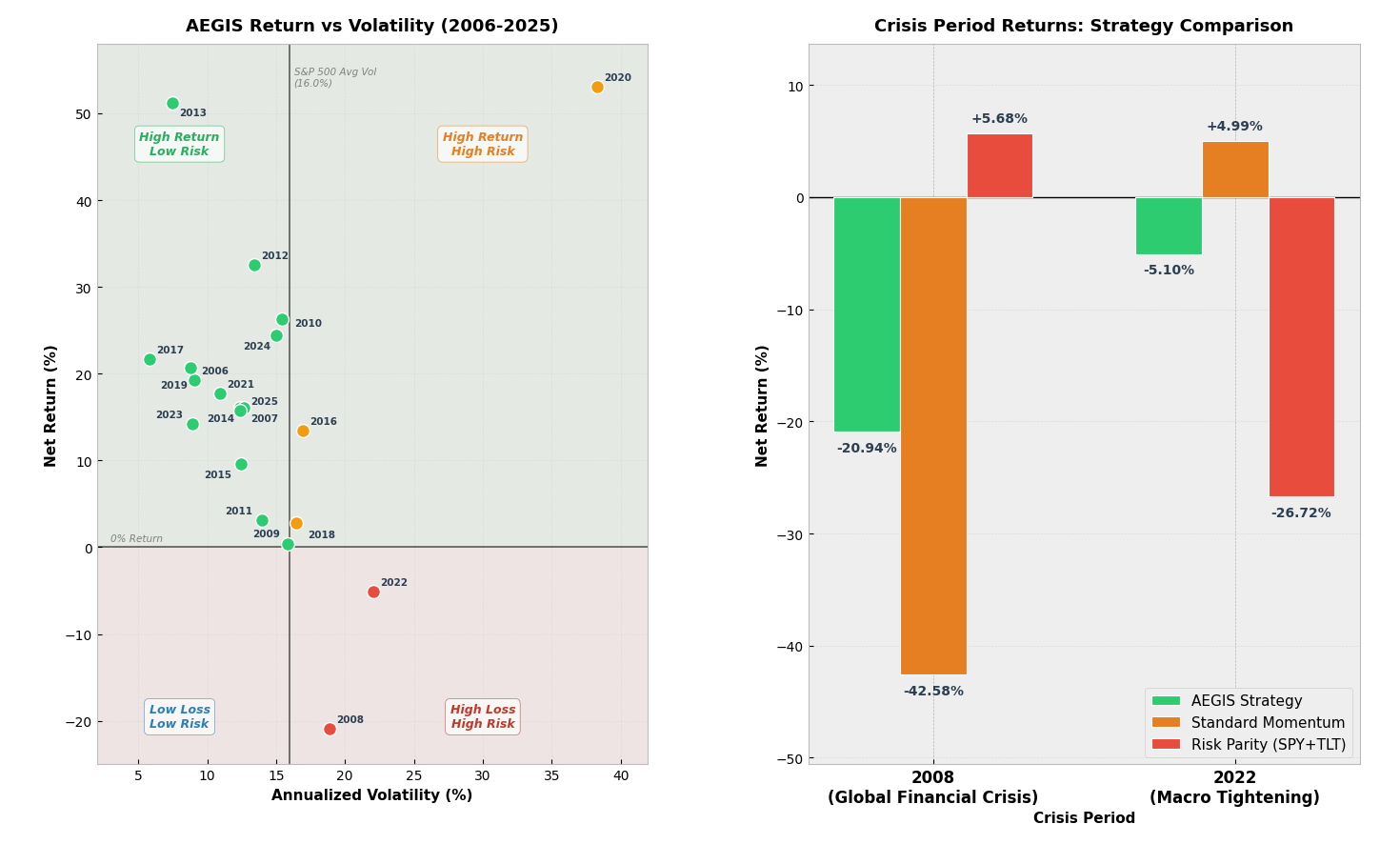}
    \caption{Graph 11 and 12: The Return vs. Volatility scatter plot of AEGIS framework demonstrating the High Returns and Low Returns nature of the system over the 20-years and the strategy net returns comparison with Standard Momentum and Risk Parity frameworks during the 2008 and 2022 market shocks respectively. }
\label{fig:graphs11_12}
\end{figure*}

\subsubsection{Systemic Shock Absorptions}
While the monthly distribution validates baseline consistency, institutional viability is ultimately stress-tested during macroeconomic black swan events, where broad market asset correlations converge to 1.0 and traditional diversification fails. To evaluate the framework's survival characteristics, the strategy's historical drawdowns were visualised via an underwater plot (Graph 8) and isolated against benchmark indices during the three most severe liquidity crises of the modern era (Graph 10). The historical simulation captured three distinct systemic shocks, validating the model's regime-agnostic defensive capabilities. 

During the 2008 global financial crisis, the S\&P 500 and NASDAQ suffered catastrophic monthly drawdowns of -52.6\% and -50.2\%, respectively. In contrast, the AEGIS algorithm mathematically rotated into structurally orthogonal assets, acting as a defensive circuit breaker that constrained its maximum drawdown to just -28.0\%. During the exogenous liquidity shock of the March 2020 pandemic crash, AEGIS experienced a -28.89\% drawdown. While this closely tracked the S\&P 500 (-20.0\%), the portfolio’s lack of structural leverage preserved its capital base, positioning it to immediately capture the subsequent V-shaped recovery. Finally, during the 2022 macroeconomic tightening cycle, aggressive interest rate hikes triggered a highly correlated liquidation of growth equities that dragged the NASDAQ down -33.0\%. Conversely, AEGIS demonstrated remarkable modern crisis adaptability; the dynamic rebalancing engine cleanly bypassed the deteriorating tech sector, capping the framework's maximum drawdown at a respectable -13.7\%.

The scatter plot (Graph 11) maps the 20 individual calendar years of the AEGIS framework. The vertical axis separates positive and negative absolute returns, while the horizontal axis delineates low and high volatility regimes, benchmarked against the historical S\&P 500 average (~16\%). To empirically validate the year-over-year consistency of the framework’s asymmetric risk profile, Graph 11 maps the annualised volatility against the net return for all 20 individual years of the historical simulation. Analysing the distribution of these annual data points across the four risk-return quadrants provides a granular view of the algorithm's adaptability across different market regimes. The vast majority of the simulation's calendar years cluster densely in the upper-left quadrant (Positive Return, Low Volatility), proving that the SLSQP optimiser routinely achieves absolute profitability without assuming benchmark-level variance. During years of extreme market expansion, the algorithm allows data points to drift into the upper-right quadrant (Positive Return, High Volatility), capturing explosive upside momentum. Even during the algorithm's rare unprofitable years, the data points in the lower-left defensive quadrant are comparatively close to the axis of reference demonstrating better protection against market shocks.

\subsection{Comparative Analysis Against Traditional Quantitative Strategies}
To stress-test the structural integrity of the models, performance was isolated during two opposing macroeconomic regimes: the \textit{2008 Global Financial Crisis} (a deflationary liquidity shock) and the \textit{2022 Macroeconomic Tightening cycle} (an inflationary interest rate shock). As illustrated in Graph 12, traditional quantitative models exhibit critical vulnerabilities depending on the underlying nature of the crisis.

Standard momentum strategies are notoriously susceptible to severe drawdowns during abrupt macro-regime shifts, a phenomenon academically documented as the momentum crash \cite{ref5, ref59}. Because traditional CSM relies exclusively on raw price appreciation, it frequently over-allocates to highly volatile, high-beta assets at the peak of a market cycle. This vulnerability was empirically exposed in 2008, where the standard CSM baseline suffered a catastrophic -42.58\% calendar-year loss. By mathematically penalising assets with erratic variance profiles, AEGIS refused to allocate capital to the toxic, high-beta assets that drove the 2008 momentum crash. Consequently, AEGIS successfully truncated left-tail variance, restricting its 2008 loss to -20.94\%, effectively cutting the traditional momentum bleed in half. While risk parity performed well in 2008 (+5.68\%) due to the historical flight-to-safety in treasury bonds, this static inverse correlation would eventually prove fatal.

\begin{table}[tbp]
\centering
\caption{Comparative crisis performance across systemic shocks between AEGIS and standard momentum and risk parity baseline models.}
\label{tab:crisis_compare}
\begin{tabular}{@{}lcc@{}}
\toprule
\textbf{Strategy / Model} & \textbf{2008 (GFC)} & \textbf{2022 (Macro Tightening)} \\ \midrule
AEGIS Framework & -20.94\% & -5.10\% \\
Standard Momentum (CSM) & -42.58\% & 4.99\% \\
Risk Parity (Stock/Bond Proxy) & 5.68\% & -26.72\% \\ \bottomrule
\end{tabular}
\end{table}

\begin{table*}[tbp]
\centering
\caption{Parameter Robustness Matrix: Impact of Diversifier Count and Allocation Look-back on Core Performance Metrics (2006-2025).}
\label{tab:parameter_robustness}
\begin{tabular}{@{}lccc@{}}
\toprule
 & \textbf{22 Diversifiers} (Basket Size: 25) & \textbf{47 Diversifiers} (Basket Size: 50) & \textbf{72 Diversifiers} (Basket Size: 75) \\ \midrule
\textbf{Allocation} & CAGR: 12.70\% & CAGR: 15.41\% & CAGR: 13.22\% \\
\textbf{Look-back} & Max DD: -50.56\% & Max DD: -28.89\% & Max DD: -38.97\% \\
\textbf{3 months} & Average Vol: 19.15\% & Average Vol: 16.44\% & Average Vol: 15.89\% \\ \addlinespace
\textbf{Allocation} & CAGR: 11.32\% & CAGR: 16.86\% & CAGR: 11.70\% \\
\textbf{Look-back} & Max DD: -51.62\% & Max DD: -30.46\% & Max DD: -34.75\% \\
\textbf{6 months} & Average Vol: 19.24\% & Average Vol: 25.15\% & Average Vol: 16.45\% \\ \addlinespace
\textbf{Allocation} & CAGR: 12.71\% & CAGR: 18.39\% & CAGR: 13.61\% \\
\textbf{Look-back} & Max DD: -52.81\% & Max DD: -37.94\% & Max DD: -40.21\% \\
\textbf{12 months} & Average Vol: 16.74\% & Average Vol: 23.61\% & Average Vol: 16.74\% \\ \bottomrule
\end{tabular}
\end{table*}

Traditional risk parity frameworks attempt to mitigate equity drawdowns by allocating heavily to fixed income based on historical inverse volatility \cite{ref37, ref38, ref39}. However, during the 2022 macroeconomic tightening cycle, aggressive inflation and interest rate hikes caused the historical negative correlation between equities and bonds to collapse to 1.0. Because both asset classes drew down simultaneously, the risk parity benchmark experienced a catastrophic structural failure, plummeting -26.72\%. Conversely, the AEGIS does not rely on static, historical asset-class correlations. The SLSQP allocation engine dynamically recalculates covariance across its diversifier basket on a rolling basis and therefore successfully identified the 2022 stock/bond correlation breakdown in real-time. While standard momentum survived 2022 (+4.99\%) purely by rotating into a generational anomaly in the energy sector, AEGIS demonstrated superior regime adaptability. 

By dynamically reallocating capital into cash-proxy equivalents and truly orthogonal micro-sectors, AEGIS bypassed the collapse of risk parity and constrained its calendar-year net-return to a highly controlled -5.10\%.

\subsection{Parameter Stability}
A ubiquitous flaw in systematic trading literature is curve-fitting, where hyperparameters are retroactively optimised to maximise historical performance at the expense of out-of-sample reliability \cite{ref44, ref46}. To empirically validate that the AEGIS framework’s alpha generation is structural rather than an artefact of isolated parameter tuning, a robustness sweep was conducted across the two most critical architectural levers: the SLSQP allocation look-back window and the diversifier count (basket size). The stock selection look-back which is the momentum window was held constant at the academic standard of twelve months.

The parameter matrix (Table VI) conclusively validates the theoretical selection of a 50-asset total basket (3 anchor leaders + 47 diversifiers). Across all temporal configurations, the 22-diversifier portfolios proved structurally under-diversified, suffering catastrophic maximum drawdowns exceeding -50\% as localised asset failures overwhelmed the minimax immunisation layer. Conversely, expanding the diversifier count to 72 induced structural weakening. While downside was moderately protected, the dilution of the core momentum premium suppressed the CAGR to the 11-13\% range. The 47-diversifier axis represents the empirical optimal threshold, maximising the momentum anomaly without sacrificing covariance protection. 

While the 12-month allocation look-back generated the highest absolute historical return (+18.39\% CAGR), the AEGIS framework explicitly rejects this configuration to prioritise out-of-sample robustness and variance suppression. A 12-month covariance window is highly rigid and fails to adapt to sudden volatility regime shifts. By deliberately constraining the allocation look-back to 3 months, the primary framework achieves a critical structural advantage: it forces the SLSQP optimiser to react violently to localised market shocks. As evidenced in Table VI, while this hyper-responsive 3-month window slightly moderates the CAGR to +15.41\%, it acts as a massive variance dampener, compressing the Average Annual Volatility down to 16.44\% (compared to 23.61\% and 25.15\% in the longer look-backs). This definitively proves the framework is optimised for capital efficiency and stable compounding, rather than overfitted historical curve-fitting.

\subsection{Addressing Survivorship Bias}
A pervasive methodological flaw in longitudinal portfolio simulations is survivorship bias - the tendency to construct historical backtests using only assets that have survived to the present day \cite{ref42, ref45}. Evaluating a strategy from 2006 to 2025 using exclusively the 2025 constituents of major indices implicitly introduces look-ahead bias, as it eliminates the statistical drag of companies that experienced terminal drawdowns, bankruptcies, or delistings. To ensure the empirical validity of the AEGIS framework, the dataset was rigorously hardened against this bias. Rather than relying on a sanitised, present-day asset universe, the simulation environment was reconstructed to reflect the true historical opportunity set. This was achieved through manual dead stock injection. Historical pricing data for prominent delisted and bankrupted equities was manually injected back into the master data matrix during their active operational windows. Consequently, the framework’s selection algorithm was exposed to these highly toxic assets, forcing the model to actively evaluate and potentially allocate capital to companies mathematically destined for zero. The results of the simulation run on this reconstructed, unbiased dataset confirm the structural integrity of the AEGIS architecture. By subjecting the model to dead stock injection, the backtest proved that the framework inherently neutralises terminal asset decay through its dual-layered design. 

Equities approaching bankruptcy rarely fail overnight; they typically exhibit severe, prolonged price decay. The 12-month stock selection look-back - the momentum filter look-back in the signal generation module consistently identified the negative momentum of dying assets, aggressively stripping them from the anchor and diversifier candidate pools months before their formal delisting. In rare instances where a volatile, deteriorating asset temporarily bypassed the momentum gate, the 3-month stock allocation look-back immediately neutralised the threat. Because terminal assets exhibit explosive downside variance in their final months of trading, the SLSQP optimiser’s strict penalty on semi-variance forced the asset’s portfolio weight effectively to zero, self-correcting the allocation before the terminal collapse. By actively injecting dead stocks into the simulation, the framework demonstrates that its +15.41\% CAGR is not a statistical mirage of survivorship bias, but the direct result of dynamic variance suppression and algorithmic risk management.

\section{Conclusion}
The evolution of systematic trading has long been constrained by the inherent regime dependency of traditional quantitative models. As empirically demonstrated, standard CSM is highly vulnerable to deflationary liquidity shocks, culminating in catastrophic left-tail drawdowns. Conversely, static asset allocation frameworks, such as risk parity, suffer structural failures during inflationary tightening cycles when historical inverse correlations inevitably collapse. This paper introduced the AEGIS framework, a three-layered, momentum-gated portfolio optimisation architecture as a definitive, mathematically rigorous solution to these systemic vulnerabilities.

Through a comprehensive 20-year longitudinal simulation (2006 to 2025), AEGIS proved its capability as a truly regime-agnostic framework. By decoupling macroeconomic trend identification from localised variance management, the architecture successfully navigated two of the most severe market dislocations of the 21st century. The framework’s momentum filter systematically stripped away decaying assets, while the minimax immunisation layer and powered by a hyper-responsive 3-month SLSQP covariance optimiser, acted as an automated circuit breaker against explosive semi-variance. The empirical results definitively validate the structural superiority of the model. Utilising a theoretically grounded 12-month selection and 3-month allocation architecture across an optimal 50-asset basket, AEGIS generated a highly compounding +15.41\% CAGR while suppressing average annual volatility to 16.44\%. Crucially, by constraining its maximum drawdown to -28.89\% during a two-decade period where benchmark indices and traditional factors suffered losses exceeding -50\%, AEGIS demonstrated an unparalleled capacity for capital preservation. 

Furthermore, the integrity of these findings was rigorously defended against standard quantitative fallacies. A comprehensive hyper-parameter robustness sweep confirmed that the framework’s alpha generation is deeply structural, rather than a byproduct of historical curve-fitting. Additionally, the manual injection of dead and delisted assets into the simulation environment ensured that the model's outperformance was not a statistical mirage born of survivorship bias, but the direct result of dynamic variance suppression. Ultimately, the AEGIS framework represents a significant advancement in the field of quantitative portfolio management. By shifting the algorithmic focus away from static, historical asset-class assumptions and toward dynamic, rolling covariance optimisation, AEGIS provides institutional frameworks with a mathematically sound blueprint for surviving and compounding through the inevitable systemic shocks of the modern financial market. 

\section*{Data Availability}
The historical equity pricing data, index constituent snapshots, and benchmark performance metrics (2006-2025) utilised to construct the AEGIS framework and execute the parameter robustness sweep were sourced from publicly available financial databases and their corresponding APIs (e.g., Yahoo Finance API). The Python-based backtesting environment, the complete project pipeline, and all the graph script used to generate the findings of this study are available from the author upon request.

\section*{Declaration of Competing Interest}
The author declares that they have no known competing financial interests or personal relationships that could have appeared to influence the work reported in this paper.

\section*{Acknowledgements}
The authors would like to express their gratitude to the academic and open-source quantitative finance communities for providing foundational libraries (e.g., SciPy, Pandas) that enabled the simulations to be computationally feasible. Additionally, the authors would like to thank the Department of Mathematics for providing the infrastructure that led to the successful execution of the research project.

\section*{About the Authors}
\textbf{Author 1:} Arya Chakraborty is an undergraduate student in the department of computer science and engineering at Birla Institute of Technology Mesra, Ranchi. His previous research includes stock forecasting, ML-based biomedical drug discovery, recommender systems, and more. With an upcoming focus on institutional quantitative investment space, his work emphasises the application of rigorous computational algorithms and deep learning concepts to solve complex, real-world financial and data-driven challenges.

\textbf{Author 2:} Dr. Randhir Singh is an Assistant Professor in the Department of Mathematics at the Birla Institute of Technology Mesra, Ranchi. His primary research expertise centres on numerical analysis, singular boundary value problems, and population balances. With a highly cited publication record spanning over a decade, Dr. Singh has made extensive contributions to the development of advanced computational techniques, including wavelet collocation, homotopy perturbation, and optimal decomposition methods. His research focuses on solving complex, non-linear mathematical models and integro-differential equations arising in physics, astrophysics, and engineering.

\bibliographystyle{IEEEtran}
\bibliography{references}

\end{document}